\documentclass[12pt,preprint]{aastex}

\begin{document}

\title{Period changes and four color light curves of the active overcontact binary V396 Monocerotis}

\author{Liu L.\altaffilmark{1,2,3,4}, Qian S.-B.\altaffilmark{1,2,3,4}, Liao
W.-P.\altaffilmark{1,2,3,4}, He J.-J.\altaffilmark{1,2,3,4}, Zhu
L.-Y.\altaffilmark{1,2,3,4}, Li L.-J.\altaffilmark{1,2,3,4} and Zhao
E.-G.\altaffilmark{1,2,3,4}} \singlespace

\altaffiltext{1}{National Astronomical Observatories/Yunnan
Observatory, Chinese Academy of Sciences, P.O. Box 110, 650011
Kunming, P.R. China (e-mail: creator\_ll.student@sina.com;
LiuL@ynao.ac.cn)}

\altaffiltext{2}{Key Laboratory for the Structure and Evolution of
Celestial Objects, Chinese Academy of Sciences, 650011 Kunming, P.
R. China}

\altaffiltext{3}{United Laboratory of Optical Astronomy, Chinese
Academy of Sciences (ULOAC), 100012 Beijing, P. R. China}

\altaffiltext{4}{Graduate School of the CAS, Beijing, P.R. China}

\begin{abstract}
This paper analyzes the first secured four color light curves of
V396 Mon using the 2003 version of the WD code. It is confirmed that
V396 Mon is a shallow W-type contact binary system with a mass ratio
$q=2.554(\pm0.004)$ and a degree of contact factor
$f=18.9\%(\pm1.2\%)$. A period investigation based on all available
data shows that the period of the system includes a long-term
decrease ($dP/dt=-8.57\times{10^{-8}}$\,days/year) and an
oscillation ($A_3=0.^{d}0160$; $T_3=42.4\,years$). They are caused
by angular momentum loss (AML) and light-time effect, respectively.
The suspect third body perhaps is a small M-type star (about 0.31
solar mass). Though some proofs show that this system has strong
magnetic activity, through analyzing we found that the Applegate
mechanism cannot explain the periodic changes. This binary is an
especially important system according to Qian's statistics of
contact binaries as its mass ratio lies near the proposed pivot
point about which the physical structure of contact binaries
supposedly oscillate.
\end{abstract}

\section{Introduction}

The light curves of V396 Monocerotis are continuous, sine-like. The
primary minima are nearly of the same depth as the secondary minima.
That is, V396 Mon has an EW-type light curve and was classified as a
W Ursae Majoris eclipsing binary in the General Catalogue of
Variable Stars (Kholopov 1985). A photographic light curve, times of
light minimum, and ephemeris were given by Wachmann (1964).
Subsequently, additional times of minima were published by Hoffmann
(1983) and the Swiss Astronomical Society (BBSAG). Then, $BV$ light
curves observed in 1999, photoelectric solutions, and a period
analysis were given by Yang \& Liu (2001). Their conclusions were
that V396 Mon is a W-type W UMa contact binary with a mass ratio of
0.402 and a cool spot presented on the secondary component which
caused asymmetry of the light curves. One year later, Gu (2004)
found the O'Connell effect (O'Connell 1951) in his observation to
become very weak compared with that in Yang \& Liu's (2001)
observation; he thought that this was an indication of starspot
activity.

Generally, magnetic activity probably leads to an alternate period
change of a close binary (e.g., Applegate 1992, Lanza et al. 1998).
However, in Yang \& Liu's (2001) period analysis they did not find
the expected variations owing to a lack of times of light minima
data. Qian has published a series papers to discuss the long-term
period variation of contact binary stars (i.e., Qian 2001a, b,
2003). His result is that this kind of variation may correlate with
the mass of the primary component ($M_{1}$), and the mass ratio of
the system ($q$). His statistic critical mass ratio q is 0.4. When
$q>0.4$, the secular period increases; contrary, $q<0.4$, the
secular period decreases. V396 Mon is important because of its mass
ratio is around 0.4. As Qian said, such systems should be unstable
and oscillate around the critical mass ratio. Therefore, V396 Mon
becomes one of the monitoring targets in our contact binaries
observation program running at Yunnan Observatory.

\section{Observations}

We carried out new CCD photometric observations of V396 Mon in
$BVRI$ bands on 2009 November 16 and 17, using the $1024 \times
1024$ PI1024 BFT camera attached to the 85-cm telescope at the
Xinglong Station of National Astronomical Observatories of Chinese
Academy of Sciences. The filter system was a standard
Johnson-Cousins-Bessel multicolor CCD photometric system built on
the primary focus (Zhou et al. 2009). The effective field of view is
$16.'5 \times 16.'5$. The integration time for each image is 20\,s.
2MASS06382732+0340193 and 2MASS06385506+0339462 were chosen as
comparison and check star, respectively. These two stars are close
to the target and their brightnesses are similar to the target. PHOT
(measure magnitudes for a list of stars) of the aperture photometry
package of IRAF was used to reduce the observed images, including a
flat-fielding correction process. Through the observation we
obtained $BVRI$ light curves. The original data are listed in Tables
1 to 4. By calculating the phase of the observations with the
equation $2455153.3614+0.^{d}39634359 \times E$, the light curves
are plotted in Figure 1. In this figure it is shown that the data of
two days joined smooth and the light variation is typical of EW
type. The magnitude difference between comparison star and check
star is a constant, denoting the authenticity for the variations of
the curves about V396 Mon. Since the light levels around the minima
are symmetric, a quadratic polynomial fitting method was used to
determine the times of minimum light by the least square method. In
all, our new epochs of light minima were listed in Table 5.

The light curves in the V band obtained by Yang and Liu (2001) in
1999 and by the authors in 2009 are plotted in Figure 2. As shown in
this figure, the light curves changed between 1999 January and 2009
November. The light curves observed in 2009 November are symmetric,
while those obtained in 1999 January are asymmetric, which exhibit a
typical O' Connell effect and show a much deeper primary minimum.
The similar phenomenon was found out by Qian et al. (2006) when they
analyzed a deep, low mass ratio overcontact binary system, AH
Cancri. They did not interpret it then. Now, we are convinced that
the changes of the light curves are caused by some cool star spots
on the surface of the components. Actually, a cool star spot or
several ones can make the primary minimum much deeper, which was
proofed by Yang and Liu's photometric solutions. Later, the spot
disappeared, which was proofed by the photometric solutions of Gu
(2004) and us. The present cool spots will alter the light curves
very much, even strongly affecting the results of the photometric
parameters.

\section{Orbital period analysis}

The first orbital period analysis of V396 Mon was given by Yang \&
Liu (2001). They collected 30 light minima and yield a corrected
ephemeris,
\begin{eqnarray}
{\rm Min.~I}&=&2451199.0715(\pm0.0006)\nonumber\\
    & &+0.^{d}3963410(\pm0.0000007)\times{E}.
\end{eqnarray}

\noindent As time passed, more and more minima of V396 Mon were
obtained by various observers. We collected all available visual,
photoelectric and CCD times of light minima, listing them in Table
6. We adopted the ephemeris $2455153.3614+0.^{d}39634498 \times E$
to modify its period, where 2455153.3614 is one of our times of
light minima and 0.39634498 is found in GCVS (Samus et al. 2004). In
calculation, the weight of visual data being set as 1 meanwhile that
of the others being set as 8. A new corrected linear ephemeris was
obtained:
\begin{eqnarray}
{\rm Min.~I}&=&2455153.3766(\pm0.0035)\nonumber\\
    & &+0.^{d}39634359(\pm0.00000010)\times{E}.
\end{eqnarray}
\noindent The $(O-C)$ values with respect to the linear ephemeris
are listed in the fifth column of Table 6. The corresponding
$(O-C)_{1}$ diagram is displayed in Figure 3.

The general $(O-C)_1$ trend of V396 Mon shown in Figure 3 indicates
the continuous period decrease. However a long-term period decrease
alone (dashed line in Figure 3) cannot describe the $(O-C)_1$ curve
very well; a period oscillation exists. Assuming that the period
oscillation is cyclic, then, based on a least-square method, a
sinusoidal term was added to a quadratic ephemeris to give a better
fit to the $(O-C)_1$ curve (solid line in Figure 3). The result is
\begin{eqnarray}
{\rm Min.~I}&=&2455153.3763(\pm0.0053) \nonumber\\
    &  & +0.^{d}39634132(\pm0.00000038)\times{E}\nonumber\\
    & &-4.65(\pm0.59)\times{10^{-11}}\times{E^{2}}\nonumber\\
    & &+0.0160(\pm0.0009)\sin[0.^{\circ}00921\times{E}\nonumber\\
    & &-65.^{\circ}5(\pm12.^{\circ}2)].
\end{eqnarray}

\noindent With the quadratic term in this equation, a secular period
increase rate is determined,
$dP/dt=-8.57\times{10^{-8}}$\,days/year.

The $(O-C)_{1}$ values with respect to the quadratic ephemeris in
Eq.(3) are shown in Figure 4. Although the visual data show large
scatter, most of the photoelectric and CCD data lie close to the
fitting line; an oscillation is presented in this figure. We have
the relation,
\begin{equation}
\omega=360^{\circ}P_{e}/T,
\end{equation}
\noindent where $P_{e}$ is the ephemeris period ($0.^{d}39634359$),
the period of the orbital period oscillation is determined to be
$T_3=42.4$\,years.

\section{Photometric solutions}

V396 Mon is an ignored but important system. Yang and Liu (2001)
determined its photometric solutions. They found the mass ratio $q$
to be 0.402 and the fill-out factor $f$ to be $5\%$, including a
dark spot on the secondary component. But one year later, Gu (2004)
found that the spot had dispersed. He derived a mass ratio of 2.937
(0.340).

To obtain initial input parameters, a q-search method with the 2003
version of the W-D program (Wilson \& Devinney, 1971, Wilson, 1990,
1994, Wilson \& Van Hamme, 2003) was used (Figure 5). We fixed q to
0.3, 0.4, 0.5 and so on, as figure 5 shows. It can be seen that
there are two lower points and the best value is around $q=2.5$
(0.4), which is very close to the photometric value $q=0.402$ (Yang
\& Liu 2001).

Throughout the solution the same value of temperature for star 1
(the star eclipsed at primary minimum) as that used by Yang \& Liu
(2001) ($T_1=6210$K) was chosen. The bolometric albedo $A_1=A_2=0.5$
(Rucinski 1969) and the values of the gravity-darkening coefficient
$g_1=g_2=0.32$ (Lucy 1967) were used, which correspond to the common
convective envelope of both components. Logarithm limb-darkening
coefficients were used, taken from Claret \& Gimenez (1990). We
adjusted the mass ratio ($q$); the orbital inclination ($i$); the
mean temperature of star 2 ($T_2$); the monochromatic luminosity of
star 1 ($L_{1}$) and the dimensionless potential of star 1
($\Omega_1=\Omega_2$, mode 3 for contact configuration). Like Gu
(2004)'s light curves, our multi-color light curves presented no
O'Connell effect (O'Connell 1951). Period analysis indicated that
the period oscillation may be caused by a light-time effect of a
tertiary component, so we tried to adjusted the parameter $l_3$ in
the W-D code. However, the numerical third light calculated by the
program tended to negative. Therefore, in our final results we set
third light equal to zero. The photometric solutions are listed in
Table 7 and the theoretical light curves computed with those
photometric elements are plotted in Figure 6. The geometrical
structure of V396 Mon is displayed in Figure 7.

\section{Discussions and conclusions}
V396 Mon is a W-type marginal contact binary with a reliable
photometric mass ratio 2.554 and a fill-factor 18.7\%. The mass
ratio $1/2.554=0.392$ is close to that of Yang and Liu (2001)'s
result 0.402, but the fill-factors of the two results are
significant difference ($18.7\%\pm1.2\%$ and $4.7\%\pm5.0\%$,
respectively). The reasons are the system's period decease, the
models simply unreliable in the sense of probable errors and the
real much changes of the system in a decade. First, long-term period
decrease cannot cause so big variations. The time scale of that is
at least several million years, which is much longer than a decade.
So it is impossible to see clear changes in systematic fashion in so
short time. Second, although the models are simply, the prevenient
practices tell us that the errors should not be so big. The third
reason is the main one. The appearance of the cool spot strongly
affects the results of the photometric parameters. More or less, two
deferent light curves should have two deferent photometric
solutions. Yang and Liu's (2001) light curves exist a star spot,
which impacts on the solutions of the internal physical parameters.
Moreover, the errors of their fill-factor is $5\%$, indicating the
relative error is $106.3\%$, much bigger than $6.3\%$ of ours. Thank
to the disappeared cool spots, which make the light curves to
restore their originally appearance. Hence, we think our results
reveal the uncovered physical parameters and are more reliable than
Yang and Liu's.

Although no spectroscopic elements have been published for this
binary system, their absolute parameters can be estimated. Assuming
that the primary components are normal, main sequence stars, we can
estimate their masses as 0.36 and 0.92$M_{\odot}$, corresponding to
the results of W-D code. Combining with the photometric solutions
and its period, we then estimated absolute parameters for the system
($R_1=0.84R_{\odot}$, $R_2=1.27R_{\odot}$; $L_1=0.947L_{\odot}$,
$L_2=2.043L_{\odot}$). The evolutionary status of the components can
be inferred from their mean densities (see for example, Mochnacki
1981, 1984). Using the following formulae (Kopal 1959),
\begin{equation}
\overline{\rho_1}=\frac{0.079}{V_1(1+q)P^2},~~\overline{\rho_2}=\frac{0.079q}{V_2(1+q)P^2},
\end{equation}
where $V_{1,2}$ are the volumes of the components using the
separation A as the unit of length, we determined the mean densities
($\rho_1$, $\rho_2$) of the two components to be 1.112
$\rho_{\odot}$ and 0.809 $\rho_{\odot}$. The corresponding
logarithms of the mean densities are 0.046 and -0.092, which are
lower than those of zero-age main sequence (ZAMS) stars of the same
spectral type, especially for the less massive components. This
indicates that the components in both systems have already moved
away from the ZAMS line to a greater or lesser extent.

The period variation of V396 Mon represented is very complex. Based
on all available photoelectric and CCD eclipse times, the period
changes of the contact binary star were discussed in the previous
section. First, the orbital period was revised as 0.39634359 days by
using the 118 visual, CCD and photoelectric timings being listed in
Table 6. Second, the general $(O-C)$ trend reveals a long-term
period decrease at a rate of
$dP/dt=-8.57\times{10^{-8}}$\,days/year. In addition, a period
oscillation ($A_3\sim0^{d}.0160$) was discovered superimposed on the
period decrease. If this period decrease is due to a conservative
mass transfer from the more massive component to the less massive
one, then with the absolute parameters derived by the present paper
and by using the well-known equation,
\begin{equation}
\frac{\dot{P}}{P}=3\frac{\dot{M_2}}{M_2}(\frac{M_2}{M_1}-1),
\end{equation}
\noindent the mass transfer rate is estimated to be,
$dM_2/dt=-4.26\times{10^{-8}}M_{\odot}/year$. The negative sign
implies that the more massive component $M_2$ is losing its mass.
The timescale of mass transfer is
$\tau\sim{M_2/\dot{M_2}}\sim2.2\times{10^{7}}$\,years which is 4
times the thermal time scale of the more massive component. However,
having considered the strong magnetic activity of the system (Gu
2004 and our discussions above), the long-term period decrease can
be reasonably explained as the results of an enhanced stellar wind
and angular momentum loss (AML). Table 8 lists some contact binary
systems that exhibit long-term period decreases.

We now address the short-term period oscillation with $T_3=42.4$
years, $A_3=0.0160$ days. There are two main ways to interpret these
observations: the Applegate mechanism and light-time effect.

The Applegate mechanism says that the cyclic period change is caused
by magnetic activity-driven variations in the quadrupole moment of
the solar-type components (e.g., Applegate 1992, Lanza et al. 1998).
According to the formula (Lanza \& Rodon\`{o} 2002;
Rovithis-Livaniou et al. 2000),
\begin{equation}
\frac{{\Delta}P}{P}=-9\frac{{\Delta}Q}{Ma^2},
\end{equation}
\begin{equation}
{\Delta}P=\sqrt{[1-{\rm{cos}}(2{\pi}P/P_3)]}\times A_3,
\end{equation}
\noindent we got the needed quadrupole moment were ${\Delta}Q_1=1.41
\times 10^{49}$ and ${\Delta}Q_2=3.61 \times 10^{49} g \cdot cm^2$ .
However, for active close binary stars, the typical values range
from $10^{51}$ to $10^{52} g \cdot cm^2$. Therefore the Applegate
mechanism probably does not describe the short-term period changes
in V396 Mon.

The most likely explanation of the period oscillation is that a
light-time effect of a tertiary component causes this phenomenon. By
using this equation:
\begin{equation}
f(m)=\frac{4\pi^{2}} {{\it
G}T_3^{2}}\times(a_{12}^{\prime}\sin{i}^{\prime})^{3},
\end{equation}
where $a_{12}^{\prime}\sin{i}^{\prime}=A_3\times{c}$ (where c is the
speed of light), the mass function from the tertiary component can
be computed with the following equation:
\begin{equation}
f(m)=\frac{(M_{3}\sin{i^{\prime}})^{3}} {(M_{1}+M_{2}+M_{3})^{2}}.
\end{equation}
Assuming the third body and the central system are coplanar, and
taking the estimated physical parameters given in the first
paragraph of this section, the mass and the orbital radius of the
suspect third companion can be estimated. The smallest mass of a
tertiary companion should be $m_3=0.31M_{\odot}$ with a septation of
14.2 AU. The luminosity of such a small M main sequence star is only
0.026$L_{\odot}$, 0.87\% of the whole system. This could explain why
we did not find third light in W-D solutions.

In summary, V396 Mon is a middle mass ratio shallow contact binary.
Although previous light curves showed evidence of spot activity, our
current observations seem to indicate a lack of any light curve
asymmetries. The periodic variation of its period changes is most
likely caused by a third body, probably a small M star. To more
accurately determine the absolute parameters of V396 Mon, a
precision radial velocity curve must be secured. If we can determine
that its true mass ratio is indeed close to 0.4 (the 0.4 mass ratio
argument of Qian), this binary system becomes an especially
important system and is worth to do a deep research in the future.

\acknowledgments{This work is partly supported by Yunnan Natural
Science Foundation (2008CD157), Chinese Natural Science Foundation
(No.10973037 and No.10903026), The Ministry of Science and
Technology of the People¡¯s Republic of China through grant
2007CB815406 and The Chinese Academy of Sciences grant No.
O8ZKY11001. New observations of the system were obtained with the
1-m telescope at Yunnan Observatory and 85-cm telescope at Xinglong
observation base.

We are especially indebted to the anonymous referee who given us
useful comments and cordial suggestions, which helped us to improve
the paper greatly.}

\begin{table*}
\begin{tiny}
\caption{The original data of V396 Mon in B band observed by 85cm
telescope at Xinglong base, National Observatory. Hel. JD 2455100+}
\begin{tabular}{rrrrrrrrrr rrrr}
\hline
 Hel.JD&$\Delta{m}$&Hel.JD&$\Delta{m}$&Hel.JD&$\Delta{m}$&Hel.JD&$\Delta{m}$&Hel.JD&$\Delta{m}$&Hel.JD&$\Delta{m}$&Hel.JD&$\Delta{m}$\\
 \hline
52.2027 &  -.064  &  52.2633  & $-.404 $ &  52.3239  & $-.256$ &  52.3845  &$  .176$ &  53.2474  &$ -.392$ &  53.3079 &$  -.278$ &  53.3684 & $  .222$ \\
52.2042 &  -.096  &  52.2648  & $-.403 $ &  52.3254  & $-.234$ &  52.3860  &$  .141$ &  53.2488  &$ -.398$ &  53.3093 &$  -.272$ &  53.3699 & $  .214$ \\
52.2056 &  -.096  &  52.2662  & $-.411 $ &  52.3268  & $-.222$ &  52.3874  &$  .125$ &  53.2502  &$ -.390$ &  53.3108 &$  -.281$ &  53.3713 & $  .223$ \\
52.2071 &  -.120  &  52.2677  & $-.416 $ &  52.3283  & $-.223$ &  52.3889  &$  .095$ &  53.2517  &$ -.408$ &  53.3122 &$  -.268$ &  53.3727 & $  .224$ \\
52.2085 &  -.143  &  52.2691  & $-.406 $ &  52.3297  & $-.211$ &  52.3903  &$  .075$ &  53.2531  &$ -.392$ &  53.3136 &$  -.257$ &  53.3742 & $  .208$ \\
52.2100 &  -.167  &  52.2705  & $-.418 $ &  52.3311  & $-.197$ &  52.3917  &$  .036$ &  53.2546  &$ -.404$ &  53.3151 &$  -.238$ &  53.3756 & $  .186$ \\
52.2114 &  -.171  &  52.2720  & $-.393 $ &  52.3326  & $-.178$ &  52.3932  &$  .034$ &  53.2560  &$ -.389$ &  53.3165 &$  -.216$ &  53.3771 & $  .158$ \\
52.2128 &  -.184  &  52.2734  & $-.416 $ &  52.3340  & $-.165$ &  52.3946  &$  .018$ &  53.2574  &$ -.406$ &  53.3180 &$  -.213$ &  53.3785 & $  .139$ \\
52.2143 &  -.200  &  52.2749  & $-.416 $ &  52.3355  & $-.130$ &  52.3961  &$ -.005$ &  53.2589  &$ -.404$ &  53.3194 &$  -.208$ &  53.3799 & $  .104$ \\
52.2157 &  -.208  &  52.2763  & $-.408 $ &  52.3369  & $-.134$ &  52.3975  &$ -.030$ &  53.2603  &$ -.403$ &  53.3208 &$  -.198$ &  53.3814 & $  .101$ \\
52.2172 &  -.226  &  52.2778  & $-.411 $ &  52.3384  & $-.109$ &  52.3990  &$ -.041$ &  53.2618  &$ -.398$ &  53.3223 &$  -.178$ &  53.3828 & $  .073$ \\
52.2186 &  -.252  &  52.2792  & $-.409 $ &  52.3398  & $-.087$ &  52.4004  &$ -.070$ &  53.2632  &$ -.409$ &  53.3237 &$  -.164$ &  53.3843 & $  .046$ \\
52.2201 &  -.225  &  52.2807  & $-.401 $ &  52.3413  & $-.091$ &  52.4018  &$ -.084$ &  53.2647  &$ -.400$ &  53.3252 &$  -.146$ &  53.3857 & $  .018$ \\
52.2215 &  -.268  &  52.2821  & $-.405 $ &  52.3427  & $-.054$ &  52.4033  &$ -.104$ &  53.2661  &$ -.400$ &  53.3266 &$  -.137$ &  53.3871 & $  .014$ \\
52.2229 &  -.261  &  52.2835  & $-.401 $ &  52.3441  & $-.032$ &  52.4047  &$ -.125$ &  53.2675  &$ -.396$ &  53.3280 &$  -.112$ &  53.3886 & $ -.023$ \\
52.2244 &  -.273  &  52.2850  & $-.416 $ &  52.3456  & $-.004$ &  52.4062  &$ -.140$ &  53.2690  &$ -.397$ &  53.3295 &$  -.089$ &  53.3900 & $ -.039$ \\
52.2258 &  -.291  &  52.2864  & $-.407 $ &  52.3470  & $ .004$ &  52.4076  &$ -.147$ &  53.2704  &$ -.400$ &  53.3309 &$  -.076$ &  53.3915 & $ -.056$ \\
52.2273 &  -.294  &  52.2879  & $-.397 $ &  52.3485  & $ .024$ &  52.4091  &$ -.174$ &  53.2719  &$ -.399$ &  53.3324 &$  -.053$ &  53.3929 & $ -.095$ \\
52.2287 &  -.291  &  52.2893  & $-.398 $ &  52.3499  & $ .059$ &  52.4105  &$ -.177$ &  53.2733  &$ -.400$ &  53.3338 &$  -.032$ &  53.3944 & $ -.098$ \\
52.2302 &  -.303  &  52.2907  & $-.403 $ &  52.3513  & $ .081$ &  52.4119  &$ -.194$ &  53.2747  &$ -.392$ &  53.3352 &$  -.004$ &  53.3958 & $ -.127$ \\
52.2316 &  -.309  &  52.2922  & $-.394 $ &  52.3528  & $ .110$ &  52.4134  &$ -.204$ &  53.2762  &$ -.393$ &  53.3367 &$   .007$ &  53.3972 & $ -.127$ \\
52.2330 &  -.312  &  52.2936  & $-.386 $ &  52.3542  & $ .136$ &  52.4148  &$ -.195$ &  53.2776  &$ -.383$ &  53.3381 &$   .018$ &  53.3987 & $ -.137$ \\
52.2345 &  -.320  &  52.2951  & $-.387 $ &  52.3557  & $ .165$ &  52.4163  &$ -.229$ &  53.2791  &$ -.387$ &  53.3396 &$   .064$ &  53.4001 & $ -.173$ \\
52.2359 &  -.328  &  52.2965  & $-.371 $ &  52.3571  & $ .179$ &  52.4177  &$ -.230$ &  53.2805  &$ -.380$ &  53.3410 &$   .089$ &  53.4016 & $ -.187$ \\
52.2374 &  -.332  &  52.2980  & $-.372 $ &  52.3586  & $ .192$ &  52.4192  &$ -.255$ &  53.2819  &$ -.387$ &  53.3425 &$   .119$ &  53.4030 & $ -.190$ \\
52.2388 &  -.333  &  52.2994  & $-.364 $ &  52.3600  & $ .222$ &  52.4206  &$ -.281$ &  53.2834  &$ -.371$ &  53.3439 &$   .147$ &  53.4044 & $ -.208$ \\
52.2403 &  -.335  &  52.3008  & $-.369 $ &  52.3614  & $ .221$ &  52.4235  &$ -.292$ &  53.2848  &$ -.377$ &  53.3453 &$   .155$ &  53.4059 & $ -.220$ \\
52.2417 &  -.357  &  52.3023  & $-.376 $ &  52.3629  & $ .237$ &  53.2257  &$ -.323$ &  53.2863  &$ -.369$ &  53.3468 &$   .165$ &  53.4073 & $ -.254$ \\
52.2431 &  -.356  &  52.3037  & $-.359 $ &  52.3643  & $ .230$ &  53.2272  &$ -.320$ &  53.2877  &$ -.366$ &  53.3482 &$   .181$ &  53.4088 & $ -.248$ \\
52.2446 &  -.361  &  52.3052  & $-.360 $ &  52.3658  & $ .251$ &  53.2286  &$ -.333$ &  53.2891  &$ -.363$ &  53.3497 &$   .202$ &  53.4102 & $ -.259$ \\
52.2460 &  -.363  &  52.3066  & $-.343 $ &  52.3672  & $ .245$ &  53.2301  &$ -.339$ &  53.2906  &$ -.348$ &  53.3511 &$   .219$ &  53.4117 & $ -.240$ \\
52.2475 &  -.367  &  52.3081  & $-.336 $ &  52.3687  & $ .254$ &  53.2315  &$ -.346$ &  53.2920  &$ -.340$ &  53.3525 &$   .208$ &  53.4131 & $ -.238$ \\
52.2489 &  -.368  &  52.3095  & $-.325 $ &  52.3701  & $ .260$ &  53.2329  &$ -.352$ &  53.2935  &$ -.351$ &  53.3540 &$   .207$ &  53.4145 & $ -.270$ \\
52.2503 &  -.381  &  52.3110  & $-.333 $ &  52.3715  & $ .238$ &  53.2344  &$ -.344$ &  53.2949  &$ -.338$ &  53.3554 &$   .222$ &  53.4160 & $ -.277$ \\
52.2518 &  -.385  &  52.3124  & $-.320 $ &  52.3730  & $ .243$ &  53.2358  &$ -.365$ &  53.2963  &$ -.332$ &  53.3569 &$   .210$ &  53.4174 & $ -.276$ \\
52.2532 &  -.385  &  52.3138  & $-.318 $ &  52.3744  & $ .239$ &  53.2373  &$ -.361$ &  53.2978  &$ -.323$ &  53.3583 &$   .233$ &  53.4189 & $ -.297$ \\
52.2547 &  -.389  &  52.3153  & $-.308 $ &  52.3759  & $ .247$ &  53.2387  &$ -.363$ &  53.2992  &$ -.332$ &  53.3598 &$   .221$ &  53.4203 & $ -.294$ \\
52.2561 &  -.396  &  52.3167  & $-.295 $ &  52.3773  & $ .244$ &  53.2401  &$ -.382$ &  53.3007  &$ -.314$ &  53.3612 &$   .211$ &          & $      $ \\
52.2576 &  -.392  &  52.3182  & $-.289 $ &  52.3788  & $ .229$ &  53.2416  &$ -.372$ &  53.3021  &$ -.314$ &  53.3626 &$   .227$ &          & $      $ \\
52.2590 &  -.392  &  52.3196  & $-.284 $ &  52.3802  & $ .228$ &  53.2430  &$ -.381$ &  53.3036  &$ -.308$ &  53.3641 &$   .231$ &          & $      $ \\
52.2605 &  -.399  &  52.3210  & $-.277 $ &  52.3816  & $ .222$ &  53.2445  &$ -.379$ &  53.3050  &$ -.285$ &  53.3655 &$   .216$ &          & $      $ \\
52.2619 &  -.397  &  52.3225  & $-.261 $ &  52.3831  & $ .194$ &  53.2459  &$ -.385$ &  53.3064  &$ -.301$ &  53.3670 &$   .230$ &          & $      $ \\
\hline
\end{tabular}\end{tiny}
\end{table*}

\begin{table*}
\begin{tiny}
\caption{The original data of V396 Mon in V band observed by 85cm
telescope at Xinglong base, National Observatory. Hel. JD 2455100+}
\begin{tabular}{llllllllll llll}
\hline
 Hel.JD&$\Delta{m}$&Hel.JD&$\Delta{m}$&Hel.JD&$\Delta{m}$&Hel.JD&$\Delta{m}$&Hel.JD&$\Delta{m}$&Hel.JD&$\Delta{m}$&Hel.JD&$\Delta{m}$\\
 \hline
52.2031 &   .102    &52.2637 &  -.224  &52.3243 &  -.070  &52.3849 &   .319 &53.2449 &  -.194  &53.3054 &  -.108 &53.3659 &   .374 \\
52.2046 &   .083    &52.2652 &  -.221  &52.3258 &  -.058  &52.3864 &   .306 &53.2463 &  -.199  &53.3068 &  -.105 &53.3674 &   .376 \\
52.2060 &   .060    &52.2666 &  -.210  &52.3272 &  -.046  &52.3878 &   .284 &53.2478 &  -.197  &53.3083 &  -.100 &53.3688 &   .390 \\
52.2075 &   .050    &52.2681 &  -.212  &52.3287 &  -.026  &52.3893 &   .263 &53.2492 &  -.199  &53.3097 &  -.095 &53.3703 &   .375 \\
52.2089 &   .027    &52.2695 &  -.221  &52.3301 &  -.028  &52.3907 &   .235 &53.2506 &  -.204  &53.3112 &  -.083 &53.3717 &   .376 \\
52.2104 &   .008    &52.2709 &  -.228  &52.3315 &  -.011  &52.3921 &   .199 &53.2521 &  -.200  &53.3126 &  -.081 &53.3731 &   .370 \\
52.2118 &   .001    &52.2724 &  -.224  &52.3330 &   .000  &52.3936 &   .191 &53.2535 &  -.208  &53.3140 &  -.067 &53.3746 &   .360 \\
52.2132 &  -.009    &52.2738 &  -.218  &52.3344 &   .007  &52.3950 &   .171 &53.2549 &  -.208  &53.3155 &  -.062 &53.3760 &   .341 \\
52.2147 &  -.024    &52.2753 &  -.226  &52.3359 &   .035  &52.3965 &   .145 &53.2564 &  -.207  &53.3169 &  -.040 &53.3775 &   .317 \\
52.2161 &  -.050    &52.2767 &  -.228  &52.3373 &   .049  &52.3979 &   .125 &53.2578 &  -.198  &53.3184 &  -.032 &53.3789 &   .294 \\
52.2176 &  -.050    &52.2782 &  -.234  &52.3388 &   .069  &52.3993 &   .110 &53.2593 &  -.217  &53.3198 &  -.028 &53.3803 &   .288 \\
52.2190 &  -.065    &52.2796 &  -.226  &52.3402 &   .088  &52.4008 &   .092 &53.2607 &  -.206  &53.3212 &  -.025 &53.3818 &   .252 \\
52.2204 &  -.083    &52.2810 &  -.213  &52.3417 &   .097  &52.4022 &   .080 &53.2622 &  -.204  &53.3227 &   .003 &53.3832 &   .230 \\
52.2219 &  -.082    &52.2825 &  -.219  &52.3431 &   .119  &52.4037 &   .069 &53.2636 &  -.218  &53.3241 &   .024 &53.3847 &   .209 \\
52.2233 &  -.081    &52.2839 &  -.213  &52.3445 &   .142  &52.4051 &   .047 &53.2651 &  -.208  &53.3256 &   .033 &53.3861 &   .176 \\
52.2248 &  -.101    &52.2854 &  -.214  &52.3460 &   .159  &52.4066 &   .038 &53.2665 &  -.218  &53.3270 &   .052 &53.3875 &   .167 \\
52.2262 &  -.090    &52.2868 &  -.222  &52.3474 &   .183  &52.4080 &   .013 &53.2679 &  -.220  &53.3284 &   .065 &53.3890 &   .136 \\
52.2277 &  -.104    &52.2883 &  -.215  &52.3489 &   .220  &52.4095 &   .016 &53.2694 &  -.212  &53.3299 &   .088 &53.3904 &   .128 \\
52.2291 &  -.121    &52.2897 &  -.218  &52.3503 &   .236  &52.4109 &   .007 &53.2708 &  -.203  &53.3313 &   .116 &53.3919 &   .097 \\
52.2306 &  -.119    &52.2911 &  -.199  &52.3517 &   .257  &52.4123 &  -.014 &53.2723 &  -.211  &53.3328 &   .140 &53.3933 &   .072 \\
52.2320 &  -.141    &52.2926 &  -.194  &52.3532 &   .286  &52.4138 &  -.007 &53.2737 &  -.206  &53.3342 &   .153 &53.3948 &   .058 \\
52.2334 &  -.137    &52.2940 &  -.199  &52.3546 &   .293  &52.4152 &  -.035 &53.2751 &  -.208  &53.3356 &   .176 &53.3962 &   .051 \\
52.2349 &  -.133    &52.2955 &  -.192  &52.3561 &   .335  &52.4167 &  -.037 &53.2766 &  -.198  &53.3371 &   .189 &53.3976 &   .025 \\
52.2363 &  -.139    &52.2969 &  -.196  &52.3575 &   .340  &52.4181 &  -.065 &53.2780 &  -.191  &53.3385 &   .230 &53.3991 &   .007 \\
52.2378 &  -.141    &52.2984 &  -.191  &52.3590 &   .363  &52.4196 &  -.067 &53.2795 &  -.207  &53.3400 &   .246 &53.4005 &   .003 \\
52.2392 &  -.164    &52.2998 &  -.187  &52.3604 &   .389  &52.4210 &  -.101 &53.2809 &  -.191  &53.3414 &   .256 &53.4020 &   .001 \\
52.2406 &  -.165    &52.3012 &  -.170  &52.3618 &   .387  &52.4224 &  -.075 &53.2823 &  -.193  &53.3429 &   .276 &53.4034 &  -.009 \\
52.2421 &  -.163    &52.3027 &  -.176  &52.3633 &   .404  &52.4239 &  -.099 &53.2838 &  -.188  &53.3443 &   .307 &53.4048 &  -.028 \\
52.2435 &  -.164    &52.3041 &  -.167  &52.3647 &   .391  &52.4253 &  -.066 &53.2852 &  -.184  &53.3457 &   .336 &53.4063 &  -.043 \\
52.2450 &  -.171    &52.3056 &  -.152  &52.3662 &   .395  &53.2261 &  -.132 &53.2867 &  -.179  &53.3472 &   .354 &53.4077 &  -.054 \\
52.2464 &  -.188    &52.3070 &  -.155  &52.3676 &   .403  &53.2276 &  -.138 &53.2881 &  -.170  &53.3486 &   .362 &53.4092 &  -.067 \\
52.2479 &  -.189    &52.3085 &  -.143  &52.3691 &   .410  &53.2290 &  -.135 &53.2895 &  -.170  &53.3501 &   .374 &53.4106 &  -.082 \\
52.2493 &  -.183    &52.3099 &  -.133  &52.3705 &   .404  &53.2305 &  -.158 &53.2910 &  -.170  &53.3515 &   .366 &53.4120 &  -.085 \\
52.2507 &  -.193    &52.3113 &  -.126  &52.3719 &   .411  &53.2319 &  -.153 &53.2924 &  -.171  &53.3529 &   .369 &53.4135 &  -.097 \\
52.2522 &  -.197    &52.3128 &  -.132  &52.3734 &   .401  &53.2333 &  -.163 &53.2938 &  -.158  &53.3544 &   .375 &53.4149 &  -.092 \\
52.2536 &  -.196    &52.3142 &  -.126  &52.3748 &   .402  &53.2348 &  -.173 &53.2953 &  -.152  &53.3558 &   .379 &53.4164 &  -.109 \\
52.2551 &  -.189    &52.3157 &  -.118  &52.3763 &   .399  &53.2362 &  -.165 &53.2967 &  -.151  &53.3573 &   .371 &53.4178 &  -.120 \\
52.2565 &  -.205    &52.3171 &  -.098  &52.3777 &   .395  &53.2377 &  -.179 &53.2982 &  -.141  &53.3587 &   .382 &53.4193 &  -.131 \\
52.2580 &  -.212    &52.3186 &  -.101  &52.3792 &   .391  &53.2391 &  -.174 &53.2996 &  -.130  &53.3601 &   .380 &53.4207 &  -.119 \\
52.2594 &  -.209    &52.3200 &  -.096  &52.3806 &   .386  &53.2405 &  -.189 &53.3011 &  -.136  &53.3616 &   .389 &        &        \\
52.2609 &  -.224    &52.3214 &  -.088  &52.3820 &   .364  &53.2420 &  -.178 &53.3025 &  -.115  &53.3630 &   .402 &        &        \\
52.2623 &  -.216    &52.3229 &  -.080  &52.3835 &   .341  &53.2434 &  -.194 &53.3039 &  -.122  &53.3645 &   .382 &        &        \\
\hline
\end{tabular}\end{tiny}
\end{table*}

\begin{table*}
\begin{tiny}
\caption{The original data of V396 Mon in R band observed by 85cm
telescope at Xinglong base, National Observatory. Hel. JD 2455100+}
\begin{tabular}{llllllllll llll}
\hline
 Hel.JD&$\Delta{m}$&Hel.JD&$\Delta{m}$&Hel.JD&$\Delta{m}$&Hel.JD&$\Delta{m}$&Hel.JD&$\Delta{m}$&Hel.JD&$\Delta{m}$&Hel.JD&$\Delta{m}$\\
 \hline
52.2035 &   .185   &52.2655 &  -.128  &52.3261 &   .040  &52.3867 &   .385 &53.2466 &  -.111  &53.3072 &  -.012 &53.3677 &   .465\\
52.2049 &   .152   &52.2670 &  -.129  &52.3276 &   .065  &52.3882 &   .358 &53.2481 &  -.101  &53.3086 &  -.012 &53.3691 &   .474\\
52.2064 &   .147   &52.2684 &  -.127  &52.3290 &   .060  &52.3896 &   .344 &53.2495 &  -.098  &53.3100 &  -.002 &53.3706 &   .444\\
52.2078 &   .127   &52.2698 &  -.123  &52.3304 &   .084  &52.3910 &   .317 &53.2510 &  -.115  &53.3115 &   .012 &53.3720 &   .463\\
52.2093 &   .114   &52.2713 &  -.124  &52.3319 &   .101  &52.3925 &   .286 &53.2524 &  -.120  &53.3129 &   .017 &53.3735 &   .452\\
52.2107 &   .098   &52.2727 &  -.116  &52.3333 &   .094  &52.3939 &   .285 &53.2539 &  -.107  &53.3144 &   .031 &53.3749 &   .445\\
52.2121 &   .077   &52.2742 &  -.122  &52.3348 &   .112  &52.3954 &   .256 &53.2553 &  -.104  &53.3158 &   .037 &53.3764 &   .402\\
52.2136 &   .079   &52.2756 &  -.124  &52.3362 &   .139  &52.3968 &   .242 &53.2567 &  -.130  &53.3173 &   .046 &53.3778 &   .392\\
52.2150 &   .071   &52.2771 &  -.129  &52.3377 &   .162  &52.3983 &   .215 &53.2582 &  -.110  &53.3187 &   .050 &53.3792 &   .378\\
52.2165 &   .058   &52.2785 &  -.119  &52.3391 &   .163  &52.3997 &   .203 &53.2596 &  -.107  &53.3201 &   .069 &53.3807 &   .355\\
52.2179 &   .042   &52.2799 &  -.119  &52.3405 &   .183  &52.4011 &   .183 &53.2611 &  -.117  &53.3216 &   .087 &53.3821 &   .344\\
52.2194 &   .028   &52.2814 &  -.120  &52.3420 &   .201  &52.4026 &   .166 &53.2625 &  -.115  &53.3230 &   .100 &53.3836 &   .312\\
52.2222 &   .020   &52.2828 &  -.115  &52.3434 &   .223  &52.4040 &   .151 &53.2639 &  -.117  &53.3245 &   .111 &53.3850 &   .289\\
52.2237 &   .004   &52.2843 &  -.110  &52.3449 &   .267  &52.4055 &   .135 &53.2654 &  -.116  &53.3259 &   .141 &53.3864 &   .269\\
52.2251 &  -.009   &52.2857 &  -.115  &52.3463 &   .259  &52.4069 &   .120 &53.2668 &  -.121  &53.3273 &   .141 &53.3879 &   .242\\
52.2266 &  -.008   &52.2872 &  -.106  &52.3478 &   .285  &52.4083 &   .112 &53.2683 &  -.104  &53.3288 &   .175 &53.3893 &   .218\\
52.2280 &  -.008   &52.2886 &  -.103  &52.3492 &   .308  &52.4098 &   .091 &53.2697 &  -.093  &53.3302 &   .172 &53.3908 &   .221\\
52.2294 &  -.023   &52.2900 &  -.102  &52.3506 &   .328  &52.4112 &   .084 &53.2711 &  -.104  &53.3317 &   .195 &53.3922 &   .192\\
52.2309 &  -.024   &52.2915 &  -.107  &52.3521 &   .354  &52.4127 &   .063 &53.2726 &  -.116  &53.3331 &   .217 &53.3936 &   .167\\
52.2323 &  -.021   &52.2929 &  -.104  &52.3535 &   .385  &52.4141 &   .056 &53.2740 &  -.111  &53.3346 &   .238 &53.3951 &   .158\\
52.2338 &  -.041   &52.2944 &  -.094  &52.3550 &   .398  &52.4156 &   .048 &53.2755 &  -.099  &53.3360 &   .256 &53.3965 &   .128\\
52.2352 &  -.060   &52.2958 &  -.101  &52.3564 &   .418  &52.4170 &   .046 &53.2769 &  -.105  &53.3374 &   .286 &53.3980 &   .115\\
52.2367 &  -.038   &52.2973 &  -.095  &52.3579 &   .459  &52.4185 &   .020 &53.2783 &  -.099  &53.3389 &   .298 &53.3994 &   .104\\
52.2381 &  -.041   &52.2987 &  -.098  &52.3593 &   .460  &52.4199 &   .016 &53.2798 &  -.097  &53.3403 &   .325 &53.4009 &   .107\\
52.2396 &  -.058   &52.3001 &  -.074  &52.3607 &   .451  &52.4213 &   .013 &53.2812 &  -.103  &53.3417 &   .367 &53.4023 &   .074\\
52.2410 &  -.055   &52.3016 &  -.077  &52.3622 &   .474  &52.4228 &   .001 &53.2827 &  -.091  &53.3432 &   .371 &53.4037 &   .071\\
52.2424 &  -.070   &52.3030 &  -.077  &52.3636 &   .480  &52.4242 &  -.012 &53.2841 &  -.083  &53.3446 &   .394 &53.4052 &   .045\\
52.2439 &  -.072   &52.3045 &  -.060  &52.3651 &   .471  &52.4257 &  -.010 &53.2855 &  -.089  &53.3461 &   .417 &53.4066 &   .054\\
52.2453 &  -.067   &52.3059 &  -.065  &52.3665 &   .485  &53.2265 &  -.041 &53.2870 &  -.080  &53.3475 &   .444 &53.4081 &   .036\\
52.2468 &  -.078   &52.3074 &  -.064  &52.3680 &   .481  &53.2279 &  -.045 &53.2884 &  -.076  &53.3490 &   .439 &53.4095 &   .037\\
52.2482 &  -.092   &52.3088 &  -.051  &52.3694 &   .481  &53.2294 &  -.051 &53.2899 &  -.078  &53.3504 &   .458 &53.4110 &   .007\\
52.2496 &  -.100   &52.3102 &  -.033  &52.3708 &   .484  &53.2308 &  -.059 &53.2913 &  -.073  &53.3518 &   .467 &53.4124 &   .025\\
52.2511 &  -.100   &52.3117 &  -.054  &52.3723 &   .482  &53.2322 &  -.053 &53.2928 &  -.063  &53.3533 &   .460 &53.4138 &   .007\\
52.2525 &  -.097   &52.3131 &  -.033  &52.3737 &   .493  &53.2337 &  -.071 &53.2942 &  -.056  &53.3547 &   .477 &53.4153 &   .009\\
52.2540 &  -.095   &52.3146 &  -.023  &52.3752 &   .501  &53.2351 &  -.070 &53.2956 &  -.054  &53.3562 &   .470 &53.4167 &  -.025\\
52.2554 &  -.098   &52.3160 &  -.010  &52.3766 &   .486  &53.2366 &  -.066 &53.2971 &  -.054  &53.3576 &   .467 &53.4182 &  -.025\\
52.2569 &  -.102   &52.3175 &  -.016  &52.3781 &   .479  &53.2380 &  -.078 &53.2985 &  -.049  &53.3591 &   .454 &53.4196 &  -.022\\
52.2583 &  -.126   &52.3189 &   .005  &52.3795 &   .477  &53.2394 &  -.084 &53.3000 &  -.032  &53.3605 &   .467 &53.4210 &  -.027\\
52.2597 &  -.107   &52.3203 &   .008  &52.3809 &   .463  &53.2409 &  -.096 &53.3014 &  -.037  &53.3619 &   .450 &        &       \\
52.2612 &  -.108   &52.3218 &   .014  &52.3824 &   .450  &53.2423 &  -.092 &53.3028 &  -.031  &53.3634 &   .451 &        &       \\
52.2626 &  -.110   &52.3232 &   .019  &52.3838 &   .424  &53.2438 &  -.090 &53.3043 &  -.028  &53.3648 &   .470 &        &       \\
52.2641 &  -.125   &52.3247 &   .036  &52.3853 &   .396  &53.2452 &  -.095 &53.3057 &  -.013  &53.3663 &   .452 &        &       \\
\hline
\end{tabular}\end{tiny}
\end{table*}

\begin{table*}
\begin{tiny}
\caption{The original data of V396 Mon in I band observed by 85cm
telescope at Xinglong base, National Observatory. Hel. JD 2455100+}
\begin{tabular}{llllllllll llll}
\hline
 Hel.JD&$\Delta{m}$&Hel.JD&$\Delta{m}$&Hel.JD&$\Delta{m}$&Hel.JD&$\Delta{m}$&Hel.JD&$\Delta{m}$&Hel.JD&$\Delta{m}$&Hel.JD&$\Delta{m}$\\
 \hline
52.2038  &  .284  &52.2644  & -.008 &52.3250  &  .146  &52.3856  &  .494  &53.2455  &  .011  &53.3075  &  .095  &53.3680  &  .555 \\
52.2052  &  .266  &52.2658  & -.008 &52.3264  &  .155  &52.3870  &  .490  &53.2470  &  .011  &53.3089  &  .102  &53.3695  &  .559 \\
52.2067  &  .255  &52.2673  & -.010 &52.3279  &  .171  &52.3885  &  .466  &53.2498  & -.002  &53.3104  &  .113  &53.3709  &  .550 \\
52.2081  &  .224  &52.2687  &  .000 &52.3293  &  .177  &52.3899  &  .429  &53.2513  & -.014  &53.3118  &  .112  &53.3723  &  .550 \\
52.2096  &  .224  &52.2702  & -.007 &52.3308  &  .183  &52.3913  &  .422  &53.2527  & -.007  &53.3132  &  .127  &53.3738  &  .546 \\
52.2110  &  .198  &52.2716  & -.015 &52.3322  &  .211  &52.3928  &  .392  &53.2542  & -.002  &53.3147  &  .135  &53.3752  &  .515 \\
52.2124  &  .179  &52.2730  & -.008 &52.3336  &  .213  &52.3942  &  .385  &53.2556  &  .005  &53.3161  &  .144  &53.3767  &  .500 \\
52.2139  &  .166  &52.2745  & -.002 &52.3351  &  .231  &52.3957  &  .362  &53.2571  & -.007  &53.3176  &  .148  &53.3781  &  .494 \\
52.2153  &  .158  &52.2759  & -.013 &52.3365  &  .240  &52.3971  &  .334  &53.2585  & -.009  &53.3190  &  .178  &53.3796  &  .459 \\
52.2168  &  .142  &52.2774  & -.015 &52.3380  &  .267  &52.3986  &  .332  &53.2599  & -.015  &53.3204  &  .175  &53.3810  &  .432 \\
52.2182  &  .149  &52.2788  & -.011 &52.3394  &  .272  &52.4000  &  .310  &53.2614  & -.010  &53.3219  &  .193  &53.3824  &  .428 \\
52.2197  &  .142  &52.2803  & -.017 &52.3409  &  .302  &52.4014  &  .292  &53.2628  & -.019  &53.3233  &  .197  &53.3839  &  .399 \\
52.2211  &  .125  &52.2817  & -.008 &52.3423  &  .308  &52.4029  &  .261  &53.2643  & -.014  &53.3248  &  .223  &53.3853  &  .379 \\
52.2225  &  .126  &52.2831  & -.013 &52.3437  &  .339  &52.4043  &  .258  &53.2657  & -.014  &53.3262  &  .230  &53.3867  &  .343 \\
52.2240  &  .115  &52.2846  & -.013 &52.3452  &  .351  &52.4058  &  .221  &53.2671  & -.009  &53.3277  &  .236  &53.3882  &  .342 \\
52.2254  &  .109  &52.2860  & -.011 &52.3466  &  .359  &52.4072  &  .224  &53.2686  & -.008  &53.3291  &  .274  &53.3896  &  .320 \\
52.2269  &  .105  &52.2875  & -.006 &52.3481  &  .386  &52.4087  &  .232  &53.2700  & -.010  &53.3305  &  .290  &53.3911  &  .282 \\
52.2283  &  .086  &52.2889  &  .010 &52.3495  &  .417  &52.4101  &  .196  &53.2715  &  .004  &53.3320  &  .302  &53.3925  &  .282 \\
52.2298  &  .087  &52.2903  &  .007 &52.3510  &  .444  &52.4115  &  .194  &53.2729  & -.006  &53.3334  &  .317  &53.3940  &  .259 \\
52.2312  &  .093  &52.2918  &  .006 &52.3524  &  .468  &52.4130  &  .176  &53.2743  & -.001  &53.3349  &  .334  &53.3954  &  .243 \\
52.2326  &  .068  &52.2932  &  .008 &52.3538  &  .470  &52.4144  &  .170  &53.2758  &  .000  &53.3363  &  .369  &53.3968  &  .221 \\
52.2341  &  .072  &52.2947  &  .012 &52.3553  &  .499  &52.4159  &  .161  &53.2772  &  .000  &53.3377  &  .377  &53.3983  &  .221 \\
52.2355  &  .079  &52.2961  &  .005 &52.3567  &  .511  &52.4173  &  .145  &53.2787  & -.005  &53.3392  &  .410  &53.3997  &  .198 \\
52.2370  &  .060  &52.2976  &  .012 &52.3582  &  .532  &52.4188  &  .123  &53.2801  &  .010  &53.3406  &  .419  &53.4012  &  .183 \\
52.2384  &  .046  &52.2990  &  .030 &52.3596  &  .578  &52.4202  &  .124  &53.2815  &  .017  &53.3421  &  .445  &53.4026  &  .168 \\
52.2399  &  .050  &52.3005  &  .039 &52.3611  &  .552  &52.4216  &  .113  &53.2830  &  .030  &53.3435  &  .463  &53.4041  &  .154 \\
52.2413  &  .050  &52.3019  &  .032 &52.3625  &  .542  &52.4231  &  .123  &53.2844  &  .021  &53.3449  &  .491  &53.4055  &  .146 \\
52.2427  &  .036  &52.3033  &  .033 &52.3639  &  .585  &52.4245  &  .091  &53.2859  &  .028  &53.3464  &  .504  &53.4069  &  .139 \\
52.2442  &  .046  &52.3048  &  .045 &52.3654  &  .550  &52.4260  &  .073  &53.2873  &  .023  &53.3478  &  .523  &53.4084  &  .139 \\
52.2456  &  .025  &52.3062  &  .050 &52.3668  &  .580  &53.2268  &  .064  &53.2887  &  .020  &53.3493  &  .544  &53.4098  &  .124 \\
52.2471  &  .018  &52.3077  &  .052 &52.3683  &  .561  &53.2282  &  .052  &53.2902  &  .035  &53.3507  &  .541  &53.4113  &  .120 \\
52.2485  &  .029  &52.3091  &  .069 &52.3697  &  .576  &53.2297  &  .052  &53.2916  &  .034  &53.3522  &  .557  &53.4127  &  .121 \\
52.2500  &  .024  &52.3105  &  .068 &52.3711  &  .581  &53.2311  &  .056  &53.2931  &  .029  &53.3536  &  .550  &53.4141  &  .108 \\
52.2514  &  .010  &52.3120  &  .067 &52.3726  &  .573  &53.2325  &  .039  &53.2945  &  .054  &53.3550  &  .553  &53.4156  &  .110 \\
52.2528  &  .013  &52.3134  &  .074 &52.3740  &  .568  &53.2340  &  .042  &53.2959  &  .049  &53.3565  &  .554  &53.4170  &  .086 \\
52.2543  &  .003  &52.3149  &  .075 &52.3755  &  .572  &53.2354  &  .035  &53.2974  &  .061  &53.3579  &  .549  &53.4185  &  .086 \\
52.2557  &  .012  &52.3163  &  .100 &52.3769  &  .582  &53.2369  &  .035  &53.2988  &  .056  &53.3594  &  .568  &53.4199  &  .055 \\
52.2572  &  .011  &52.3178  &  .095 &52.3784  &  .565  &53.2383  &  .017  &53.3003  &  .058  &53.3608  &  .547  &53.4214  &  .053 \\
52.2586  &  .003  &52.3192  &  .099 &52.3798  &  .561  &53.2397  &  .024  &53.3017  &  .064  &53.3622  &  .558  &53.4228  &  .065 \\
52.2601  &  .003  &52.3207  &  .113 &52.3813  &  .560  &53.2412  &  .025  &53.3031  &  .061  &53.3637  &  .546  &         &       \\
52.2615  &  .002  &52.3221  &  .125 &52.3827  &  .541  &53.2426  &  .012  &53.3046  &  .061  &53.3651  &  .578  &         &       \\
52.2629  &  .000  &52.3235  &  .137 &52.3841  &  .524  &53.2441  &  .009  &53.3060  &  .088  &53.3666  &  .567  &         &       \\
\hline
\end{tabular}\end{tiny}
\end{table*}

\begin{table*}
\begin{minipage}{12cm}
\caption{New CCD times of light minima for V396 Mon.}
\begin{tabular}{llllll}\hline\hline
 JD (Hel.)    & Error (days) & Method & Min. & Filters & Telescope\\
 \hline
2452705.1581 & $\pm0.0005$ & CCD& I      & V & 1m    \\
2452944.3501 & $\pm0.0011$ & CCD& II     & V & 1m    \\
2452944.3524 & $\pm0.0003$ & CCD& II     & B & 1m    \\
2452945.3409 & $\pm0.0006$ & CCD& I      & V & 1m    \\
2452945.3394 & $\pm0.0008$ & CCD& I      & B & 1m    \\
2454917.14503 & $\pm0.00170$ & CCD& II   & I & 1m    \\
2455152.37049 & $\pm0.00015$ & CCD& II   & B & 85cm    \\
2455152.37072 & $\pm0.00020$ & CCD& II   & V & 85cm    \\
2455152.37035 & $\pm0.00016$ & CCD& II   & R & 85cm    \\
2455152.37052 & $\pm0.00014$ & CCD& II   & I & 85cm    \\
2455153.36162 & $\pm0.00018$ & CCD& I    & B & 85cm    \\
2455153.36139 & $\pm0.00016$ & CCD& I    & V & 85cm    \\
2455153.36151 & $\pm0.00018$ & CCD& I    & R & 85cm    \\
2455153.36122 & $\pm0.00017$ & CCD& I    & I & 85cm    \\
\hline\hline
\end{tabular}
\begin{flushleft}
\item 1m denotes the 1 meter R-C reflect telescope in Yunnan Observatory. 85cm
denotes the 85 cm reflect telescope in Xinglong Observation base.
\end{flushleft}
\end{minipage}
\end{table*}

\begin{deluxetable}{lllrrrl}
\tablewidth{0pc} \tabletypesize{\footnotesize} \tablecaption{Times
of light minima of V396 MON.}
\tablehead{ \colhead{Hel.JD}  & \colhead{Type}  & \colhead{Method}
&\colhead{E}  & \colhead{$(O-C)_1$}& \colhead{$(O-C)_2$} &
\colhead{Reference} }
\tiny \startdata

29691.435   &  p  &   pg  &  -64242     &  $  -0.03690 $ & $   0.00937   $ &  Wachmann 1964              \\
29696.395   &  s  &   pg  &  -64229.5   &  $  -0.03119 $ & $   0.01503   $ &  Wachmann 1964              \\
30021.405   &  s  &   pg  &  -63409.5   &  $  -0.02293 $ & $   0.02028   $ &  Wachmann 1964              \\
30024.360   &  p  &   pg  &  -63402     &  $  -0.04051 $ & $   0.00268   $ &  Wachmann 1964              \\
30025.360   &  s  &   pg  &  -63399.5   &  $  -0.03137 $ & $   0.01181   $ &  Wachmann 1964              \\
30026.350   &  p  &   pg  &  -63397     &  $  -0.03223 $ & $   0.01094   $ &  Wachmann 1964              \\
30069.375   &  s  &   pg  &  -63288.5   &  $  -0.01051 $ & $   0.03227   $ &  Wachmann 1964              \\
30072.335   &  p  &   pg  &  -63281     &  $  -0.02308 $ & $   0.01966   $ &  Wachmann 1964              \\
30373.340   &  s  &   pg  &  -62521.5   &  $  -0.04104 $ & $   -0.00100  $ &  Wachmann 1964              \\
31142.445   &  p  &   pg  &  -60581     &  $  -0.04077 $ & $   -0.00743  $ &  Wachmann 1964              \\
31144.496   &  p  &   pg  &  -60576     &  $  0.02852  $ & $   0.06183   $ &  Wachmann 1964              \\
31803.575   &  p  &   pg  &  -58913     &  $  -0.01187 $ & $   0.01598   $ &  Wachmann 1964              \\
31845.415   &  s  &   pg  &  -58807.5   &  $  0.01388  $ & $   0.04139   $ &  Wachmann 1964              \\
32233.410   &  s  &   pg  &  -57828.5   &  $  -0.01149 $ & $   0.01293   $ &  Wachmann 1964              \\
33220.515   &  p  &   pg  &  -55338     &  $  -0.00019 $ & $   0.01678   $ &  Wachmann 1964              \\
33294.425   &  s  &   pg  &  -55151.5   &  $  -0.00827 $ & $   0.00817   $ &  Wachmann 1964              \\
33685.430   &  p  &   pg  &  -54165     &  $  0.00379  $ & $   0.01745   $ &  Wachmann 1964              \\
33709.385   &  s  &   pg  &  -54104.5   &  $  -0.02000 $ & $   -0.00650  $ &  Wachmann 1964              \\
34085.325   &  p  &   pg  &  -53156     &  $  -0.01189 $ & $   -0.00097  $ &  Wachmann 1964              \\
34087.305   &  p  &   pg  &  -53151     &  $  -0.01361 $ & $   -0.00270  $ &  Wachmann 1964              \\
34748.400   &  p  &   pg  &  -51483     &  $  -0.01971 $ & $   -0.01313  $ &  Wachmann 1964              \\
34769.400   &  p  &   pg  &  -51430     &  $  -0.02592 $ & $   -0.01947  $ &  Wachmann 1964              \\
34769.417   &  p  &   pg  &  -51430     &  $  -0.00842 $ & $   -0.00197  $ &  Samus et al. 2004          \\
34771.400   &  p  &   pg  &  -51425     &  $  -0.00764 $ & $   -0.00120  $ &  Wachmann 1964              \\
34773.375   &  p  &   pg  &  -51420     &  $  -0.01436 $ & $   -0.00793  $ &  Wachmann 1964              \\
34775.360   &  p  &   pg  &  -51415     &  $  -0.01108 $ & $   -0.00466  $ &  Wachmann 1964              \\
34776.345   &  s  &   pg  &  -51412.5   &  $  -0.01693 $ & $   -0.01053  $ &  Wachmann 1964              \\
34780.315   &  s  &   pg  &  -51402.5   &  $  -0.01037 $ & $   -0.00399  $ &  Wachmann 1964              \\
35129.495   &  s  &   pg  &  -50521.5   &  $  -0.00907 $ & $   -0.00486  $ &  Wachmann 1964              \\
35131.480   &  s  &   pg  &  -50516.5   &  $  -0.00579 $ & $   -0.00159  $ &  Wachmann 1964              \\
35160.415   &  s  &   pg  &  -50443.5   &  $  -0.00387 $ & $   0.00014   $ &  Wachmann 1964              \\
35161.405   &  p  &   pg  &  -50441     &  $  -0.00473 $ & $   -0.00072  $ &  Wachmann 1964              \\
35163.385   &  p  &   pg  &  -50436     &  $  -0.00645 $ & $   -0.00245  $ &  Wachmann 1964              \\
35164.390   &  s  &   pg  &  -50433.5   &  $  0.00769  $ & $   0.01168   $ &  Wachmann 1964              \\
35165.365   &  p  &   pg  &  -50431     &  $  -0.00816 $ & $   -0.00418  $ &  Wachmann 1964              \\
35184.390   &  p  &   pg  &  -50383     &  $  -0.00766 $ & $   -0.00379  $ &  Wachmann 1964              \\
35185.375   &  s  &   pg  &  -50380.5   &  $  -0.01352 $ & $   -0.00965  $ &  Wachmann 1964              \\
35186.370   &  p  &   pg  &  -50378     &  $  -0.00937 $ & $   -0.00552  $ &  Wachmann 1964              \\
35399.595   &  p  &   pg  &  -49840     &  $  -0.01722 $ & $   -0.01465  $ &  Wachmann 1964              \\
35459.460   &  p  &   pg  &  -49689     &  $  -0.00011 $ & $   0.00210   $ &  Wachmann 1964              \\
36214.475   &  p  &   pg  &  -47784     &  $  -0.01964 $ & $   -0.02173  $ &  Wachmann 1964              \\
36983.415   &  p  &   pg  &  -45844     &  $  0.01381  $ & $   0.00766   $ &  Wachmann 1964              \\
37693.455   &  s  &   pg  &  -44052.5   &  $  0.00427  $ & $   -0.00528  $ &  Wachmann 1964              \\
37694.440   &  p  &   pg  &  -44050     &  $  -0.00159 $ & $   -0.01114  $ &  Wachmann 1964              \\
37695.435   &  s  &   pg  &  -44047.5   &  $  0.00255  $ & $   -0.00701  $ &  Wachmann 1964              \\
45022.476   &  p  &   pg  &  -25561     &  $  0.03785  $ & $   0.01044   $ &  IBVS No.2344     \\
46004.632   &  p  &   vis &  -23083     &  $  0.05444  $ & $   0.02705   $ &  BBSAG No.74        \\
46039.503   &  p  &   vis &  -22995     &  $  0.04721  $ & $   0.01983   $ &  BBSAG No.75        \\
46350.643   &  p  &   vis &  -22210     &  $  0.05749  $ & $   0.03025   $ &  BBSAG No.78        \\
46406.512   &  p  &   vis &  -22069     &  $  0.04205  $ & $   0.01483   $ &  BRNO No.27      \\
46412.451   &  p  &   vis &  -22054     &  $  0.03589  $ & $   0.00868   $ &  BBSAG No.79        \\
46744.594   &  p  &   vis &  -21216     &  $  0.04297  $ & $   0.01598   $ &  BBSAG No.82        \\
46877.370   &  p  &   vis &  -20881     &  $  0.04387  $ & $   0.01698   $ &  BBSAG No.93        \\
47068.604   &  s  &   vis &  -20398.5   &  $  0.04209  $ & $   0.01537   $ &  BBSAG No.85        \\
47170.465   &  s  &   vis &  -20141.5   &  $  0.04279  $ & $   0.01617   $ &  BBSAG No.87        \\
47531.335   &  p  &   vis &  -19231     &  $  0.04195  $ & $   0.01573   $ &  BBSAG No.90        \\
47562.420   &  s  &   vis &  -19152.5   &  $  0.01398  $ & $   -0.01219  $ &  BBSAG No.91        \\
47565.412   &  p  &   vis &  -19145     &  $  0.03340  $ & $   0.00723   $ &  BRNO No.30      \\
47565.414   &  p  &   vis &  -19145     &  $  0.03540  $ & $   0.00923   $ &  BRNO No.30            \\
47801.635   &  p  &   vis &  -18549     &  $  0.03563  $ & $   0.00976   $ &  BBSAG No.92        \\
47838.701   &  s  &   vis &  -18455.5   &  $  0.04350  $ & $   0.01769   $ &  BBSAG No.93        \\
47840.667   &  s  &   vis &  -18450.5   &  $  0.02778  $ & $   0.00197   $ &  BBSAG No.93        \\
47842.661   &  s  &   vis &  -18445.5   &  $  0.04006  $ & $   0.01426   $ &  BBSAG No.93        \\
47859.503   &  p  &   vis &  -18403     &  $  0.03746  $ & $   0.01168   $ &  BBSAG No.93        \\
47880.505   &  p  &   vis &  -18350     &  $  0.03325  $ & $   0.00750   $ &  BBSAG No.93        \\
47885.464   &  s  &   vis &  -18337.5   &  $  0.03796  $ & $   0.01221   $ &  BBSAG No.93        \\
47946.298   &  p  &   vis &  -18184     &  $  0.03322  $ & $   0.00756   $ &  BAV No.56                \\
47947.288   &  s  &   vis &  -18181.5   &  $  0.03236  $ & $   0.00670   $ &  BAV No.56                \\
47947.295   &  s  &   vis &  -18181.5   &  $  0.03936  $ & $   0.01370   $ &  BBSAG No.94        \\
47947.484   &  p  &   vis &  -18181     &  $  0.03019  $ & $   0.00453   $ &  BAV No.56                \\
48153.587   &  p  &   vis &  -17661     &  $  0.03452  $ & $   0.00918   $ &  BBSAG No.96        \\
48265.366   &  p  &   vis &  -17379     &  $  0.04463  $ & $   0.01947   $ &  BBSAG No.97        \\
48934.593   &  s  &   vis &  -15690.5   &  $  0.04548  $ & $   0.02156   $ &  LBBSAG No.102       \\
49009.492   &  s  &   vis &  -15501.5   &  $  0.03555  $ & $   0.01178   $ &  BBSAG No.103       \\
50832.2573  &  s  &   ccd &  -10902.5   &  $  0.01669  $ & $   -0.00226  $ &  IBVS No.4888         \\
50841.3736  &  s  &   ccd &  -10879.5   &  $  0.01709  $ & $   -0.00183  $ &  IBVS No.4888         \\
50852.6602  &  p  &   ccd &  -10851     &  $  0.00790  $ & $   -0.01099  $ &  BBSAG No.117    \\
50872.2967  &  s  &   vis &  -10801.5   &  $  0.02539  $ & $   0.00656   $ &  BRNO No.32         \\
51129.5117  &  s  &   ccd &  -10152.5   &  $  0.01341  $ & $   -0.00458  $ &  BRNO No.32           \\
51199.0712  &  p  &   pe  &  -9977      &  $  0.01461  $ & $   -0.00314  $ &  Yang \& Liu 2001              \\
51199.2689  &  s  &   pe  &  -9976.5    &  $  0.01413  $ & $   -0.00361  $ &  Yang \& Liu 2001              \\
51200.0618  &  s  &   pe  &  -9974.5    &  $  0.01435  $ & $   -0.00340  $ &  Yang \& Liu 2001              \\
51200.2596  &  p  &   pe  &  -9974      &  $  0.01398  $ & $   -0.00377  $ &  Yang \& Liu 2001              \\
51201.2507  &  s  &   pe  &  -9971.5    &  $  0.01422  $ & $   -0.00352  $ &  Yang \& Liu 2001              \\
51455.8920  &  p  &   ccd &  -9329      &  $  0.00476  $ & $   -0.01209  $ &  Paschke Anton$^*$          \\
51841.5354  &  p  &   ccd &  -8356      &  $  0.00585  $ & $   -0.00959  $ &  IBVS No.5287           \\
51876.6169  &  s  &   ccd &  -8267.5    &  $  0.01095  $ & $   -0.00437  $ &  IBVS No.5583       \\
51952.5107  &  p  &   ccd &  -8076      &  $  0.00495  $ & $   -0.01007  $ &  BBSAG No.124    \\
52209.5375  &  s  &   ccd &  -7427.5    &  $  0.00293  $ & $   -0.01108  $ &  BRNO No.34             \\
52321.3059  &  s  &   ccd &  -7145.5    &  $  0.00244  $ & $   -0.01112  $ &  BRNO No.34           \\
52338.3479  &  s  &   ccd &  -7102.5    &  $  0.00167  $ & $   -0.01183  $ &  BBSAG No.127    \\
52602.9056  &  p  &   ccd &  -6435      &  $  0.00002  $ & $   -0.01237  $ &  IBVS No.5378        \\
52689.3079  &  p  &   ccd &  -6217      &  $  -0.00058 $ & $   -0.01261  $ &  IBVS No.5438   \\
52705.1581  &  p  &   ccd &  -6177      &  $  -0.00412 $ & $   -0.01608  $ &  Present paper                        \\
52944.3513  &  s  &   ccd &  -5573.5    &  $  -0.00428 $ & $   -0.01520  $ &  Present paper                        \\
52945.3402  &  p  &   ccd &  -5571      &  $  -0.00623 $ & $   -0.01715  $ &  Present paper                        \\
52973.8783  &  p  &   ccd &  -5499      &  $  -0.00487 $ & $   -0.01566  $ &  IBVS No.5493    \\
52981.4090  &  p  &   pe  &  -5480      &  $  -0.00470 $ & $   -0.01546  $ &  IBVS No.5676     \\
53051.5611  &  p  &   pe  &  -5303      &  $  -0.00542 $ & $   -0.01586  $ &  IBVS No.5603        \\
53055.1282  &  p  &   ccd &  -5294      &  $  -0.00541 $ & $   -0.01584  $ &  IBVS No.5592          \\
53082.2785  &  s  &   pe  &  -5225.5    &  $  -0.00464 $ & $   -0.01495  $ &  IBVS No.5583       \\
53375.5680  &  s  &   vis &  -4485.5    &  $  -0.00940 $ & $   -0.01835  $ &  OEJV No.03         \\
53407.4759  &  p  &   ccd &  -4405      &  $  -0.00716 $ & $   -0.01596  $ &  IBVS No.5741       \\
53409.4576  &  p  &   ccd &  -4400      &  $  -0.00717 $ & $   -0.01597  $ &  IBVS No.5741       \\
53632.5958  &  p  &   ccd &  -3837      &  $  -0.01041 $ & $   -0.01815  $ &  BRNO No.34              \\
53672.6266  &  p  &   ccd &  -3736      &  $  -0.01031 $ & $   -0.01785  $ &  IBVS No.5731        \\
54154.3830  &  s  &   ccd &  -2520.5    &  $  -0.00954 $ & $   -0.01467  $ &  IBVS No.5802      \\
54494.6405  &  p  &   pe  &  -1662      &  $  -0.01301 $ & $   -0.01636  $ &  IBVS No.5870      \\
54505.3422  &  p  &   ccd &  -1635      &  $  -0.01259 $ & $   -0.01588  $ &  IBVS No.5918     \\
54512.6751  &  s  &   ccd &  -1616.5    &  $  -0.01205 $ & $   -0.01530  $ &  IBVS No.5875    \\
54840.4509  &  s  &   ccd &  -789.5     &  $  -0.01239 $ & $   -0.01385  $ &  IBVS No.5918      \\
54841.4409  &  p  &   pe  &  -787       &  $  -0.01325 $ & $   -0.01471  $ &  IBVS No.5918      \\
54874.7337  &  p  &   pe  &  -703       &  $  -0.01331 $ & $   -0.01458  $ &  IBVS No.5894    \\
54917.14338 &  p  &   ccd &  -596       &  $  -0.01239 $ & $   -0.01343  $ &  Present paper                        \\
55135.32692 &  s  &   ccd &  -45.5      &  $  -0.01600 $ & $   -0.01580  $ &  Present paper                        \\
55138.30024 &  p  &   ccd &  -38        &  $  -0.01526 $ & $   -0.01504  $ &  Present paper                        \\
55152.37052 &  p  &   ccd &  -2.5       &  $  -0.01517 $ & $   -0.01488  $ &  Present paper                        \\
55153.36144 &  s  &   ccd &  0          &  $  -0.01511 $ & $   -0.01481  $ &  Present paper                        \\

\enddata
\begin{flushleft}
\item $^*$ The data from O-C gateway, http://var.astro.cz/ocgate/ocgate.php?star=v396+mon\&lang=en
\end{flushleft}
\end{deluxetable}

\begin{table}
\caption{Photometric solutions for V396 Mon.}
\begin{tabular}{lcl}
\hline
Parameters              &  Photometric elements  &  errors\\
\hline
$g_1=g_2$               &     0.32               & assumed\\
$A_1=A_2$               &     0.50               & assumed\\
$x_{1bolo}=x_{2bolo}$   &     0.644              & assumed\\
$y_{1bolo}=y_{2bolo}$   &     0.231              & assumed\\
$x_{1B}=x_{2B}$         &     0.817              & assumed\\
$y_{1B}=y_{2B}$         &     0.215              & assumed\\
$x_{1V}=x_{2V}$         &     0.728              & assumed\\
$y_{1V}=y_{2V}$         &     0.269              & assumed\\
$x_{1R}=x_{2R}$         &     0.635              & assumed\\
$y_{1R}=y_{2R}$         &     0.276              & assumed\\
$x_{1I}=x_{2I}$         &     0.543              & assumed\\
$y_{1I}=y_{2I}$         &     0.263              & assumed\\
$T_1$                   &     6210K              & assumed\\
$q$                     &     2.554               & $\pm0.004$\\
$\Omega_{in}$           &     6.0187             & --     \\
$\Omega_{out}$          &     5.4076             & --     \\
$T_2$                   &     6121K              & $\pm3$K \\
$i$                     &     $89.^{\circ}654$ & $\pm0.863$ \\
$L_1/(L_1+L_2)(B)$  &     0.3194             & $\pm0.0007$ \\
$L_1/(L_1+L_2)(V)$  &     0.3146             & $\pm0.0006$ \\
$L_1/(L_1+L_2)(R)$  &     0.3119             & $\pm0.0164$ \\
$L_1/(L_1+L_2)(I)$  &     0.3099             & $\pm0.0005$ \\
$\Omega_1=\Omega_2$     &     5.9032             & $\pm0.0074$\\
$r_1(pole)$             &     0.2898             & $\pm0.0007$\\
$r_1(side)$             &     0.3034             & $\pm0.0008$\\
$r_1(back)$             &     0.3428             & $\pm0.0015$\\
$r_2(pole)$             &     0.4429             & $\pm0.0006$\\
$r_2(side)$             &     0.4750             & $\pm0.0008$\\
$r_2(back)$             &     0.5049             & $\pm0.0010$\\
$f$                     &     $18.9\,\%$         & $\pm1.2\,\%$\\
$\sum{(O-C)_i^2}$ &   0.000005618                         \\
\hline
\end{tabular}
\end{table}

\begin{table}
\caption{A sample of shallow contact binaries with long-term period
decrease.}
\begin{tiny}
\begin{tabular}{lcccccccccc}
\hline
Name & Period (d) & $dP/dt$ (d/yr)& i($^{\circ}$) & $M_1$($M_{\odot}$)& $M_2$($M_{\odot}$)& $q_{sp}$& $q_{ph}$ & $f$(\%)& Subtype & Spectrum \\
\hline
V417~Aql  & 0.3703 & $ -5.5\times10^{-8}  $ & 84.5  & 1.395  & 0.505 &  0.362 &  0.368 &  19  & W &  G2V    \\
SS~Ari  & 0.406  & $ -4.03\times10^{-7} $ & 75.3  & 1.343  & 0.406 &  0.302 &  0.295 &  13  & W &  G0V    \\
TY~Boo  & 0.3171 & $ -2.99\times10^{-8} $ & 76.6  & 0.93   & 0.4   &  0.437 &  0.466 &  10   & W &  G5V    \\
RW~Com  & 0.2373 & $ -4.1\times10^{-9} $ & 75.2  & 0.92   & 0.31  &  0.337 &  0.343 &  17   & W &  K0V    \\
CC~Com  & 0.2211 & $ -4.39\times10^{-8} $ & 90    & 0.69   & 0.36  &  0.522 &  0.518 &  20  & W &  K5V    \\
BI~CVn  & 0.3842 & $ -1.51\times10^{-7} $ & 69.2  & 1.646  & 0.679 &  0.413 &  0.865 &  17  & A &  F2V    \\
V1073~Cyg  & 0.7859 & $ -9.20\times10^{-6}  $ & 68.4  & 1.6    & 0.51  &  0.319 &  0.32  &  4   & A &  F2V    \\
FT~Lup  & 0.4701 & $ -1.7\times10^{-7}   $ & 84.7  &        &       &  0.43  &  0.45  &  12  &   &  F0+K2V \\
U~Peg  & 0.3748 & $ -2.1\times10^{-8}  $ & 77.5  & 1.149  & 0.379 &  0.33  &  0.331 &  9   & W &  G2V    \\
AU~Ser  & 0.3865 & $ -5.2\times10^{-8}  $ & 80.6  & 0.921  & 0.646 &  0.701 &  0.7   &  9   & A &  K0V    \\
AH~Tau  & 0.3327 & $ -6.98\times10^{-8} $ &       &        &       &        &  0.502 &  9   &   &  G1V    \\
BM~UMa  & 0.2712 & $ -7.49\times10^{-8} $ & 89.5  &        &       &        &  0.54  &  17  &   &  K0V    \\
\hline
\end{tabular}\end{tiny}
\end{table}

\begin{figure}
\begin{center}
\includegraphics[angle=0,scale=1.2 ]{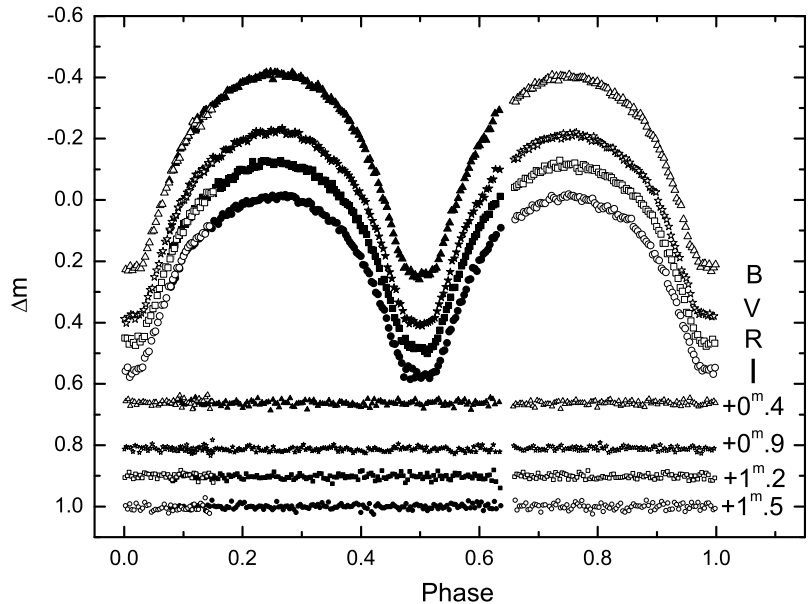}
\caption{Observed multiple-color light curves in $BVRI$ bands for
V396 Mon. Solid symbols are denote the data got in 2009-11-16; open
symbols are denote the data got in 2009-11-17. }
\end{center}
\end{figure}

\begin{figure}
\begin{center}
\includegraphics[angle=0,scale=1.2 ]{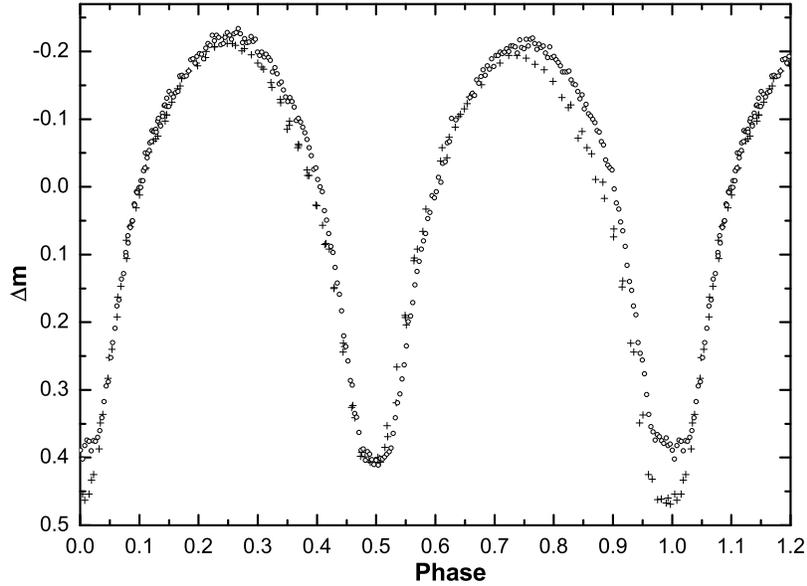}
\caption{Comparison between the V light curves obtained by Yang \&
Liu (2001) in 1999 (crosses) and by us in 2009 (circles). The change
of the light curve from phase 0.70P to 0.05P is clearly seen.}
\end{center}
\end{figure}

\begin{figure}
\begin{center}
\includegraphics[angle=0,scale=1.2 ]{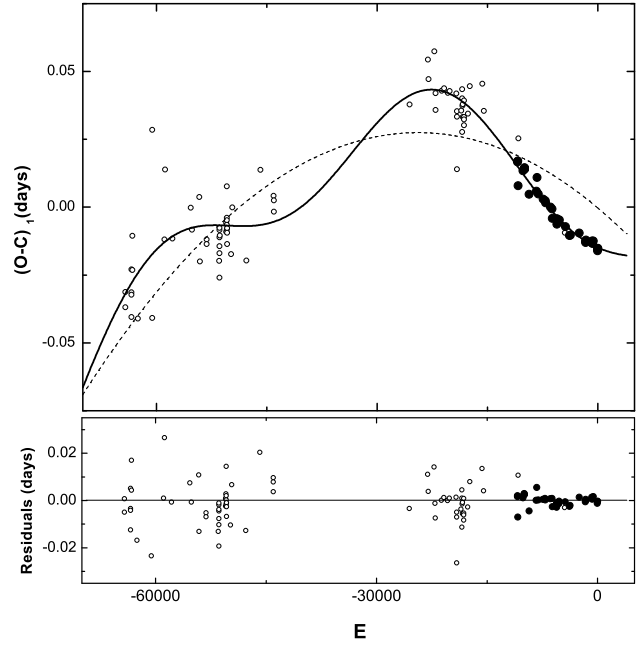}
\caption{ The $(O-C)$ diagram of V396 Mon formed by all available
measurements. The $(O-C)_1$ values were computed by using a newly
determined linear ephemeris (Eq.1). Solid circles refer to the pe
and CCD data and open ones to the visual and pg data. The dashed
line represents the quadratic fit; the solid line represents
quadratic fit superimposed on a cyclic variation (Eq.3). The lower
panel plots the residuals for Equation 3.}
\end{center}
\end{figure}

\begin{figure}
\begin{center}
\includegraphics[angle=0,scale=1 ]{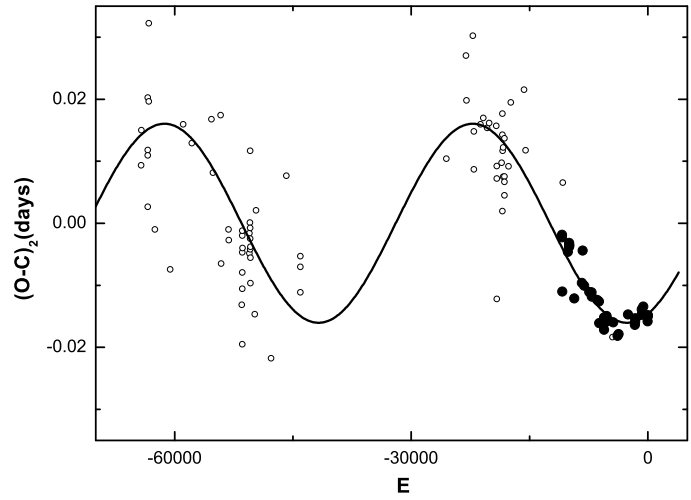}
\caption{$(O-C)_{2}$ values for V396 Mon with respect to the
quadratic ephemeris in Eq.(3). The symbols are the same as figure 3.
The solid line represents the theoretical orbit of an assumed third
body.}
\end{center}
\end{figure}

\begin{figure}
\begin{center}
\includegraphics[angle=0,scale=1 ]{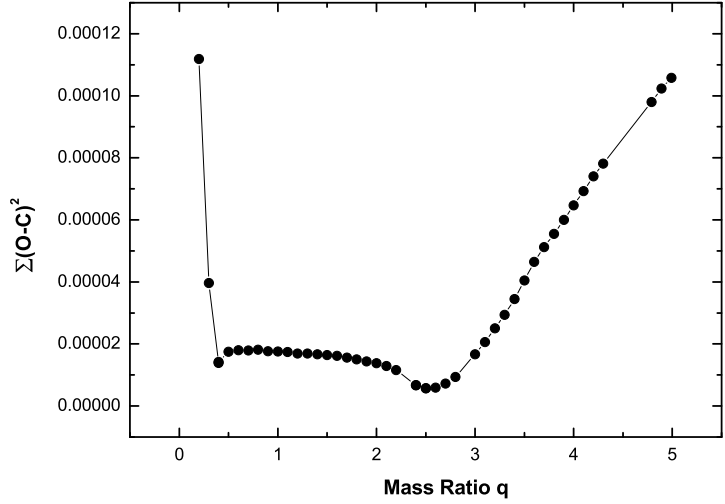}
\caption{ The relation between the mass ratio q and the sum of the
squares of the residuals $\Sigma$ for V396 Mon. }
\end{center}
\end{figure}

\begin{figure}
\begin{center}
\includegraphics[angle=0,scale=1 ]{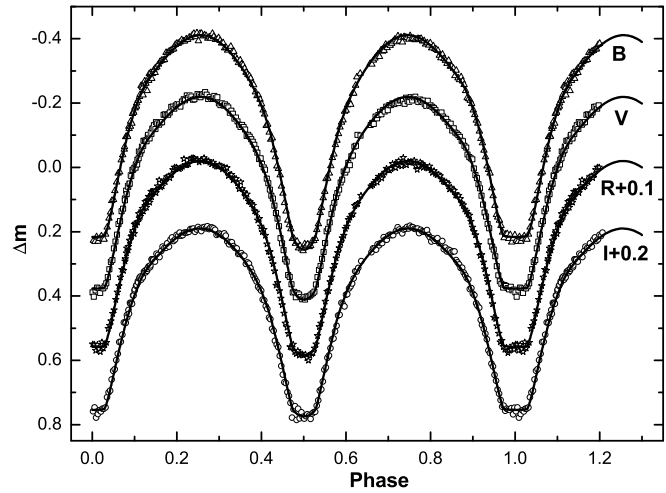}
\caption{Observed (circles) and theoretical (solid curves) light
curves in the $BVRI$ bands for V396 Mon.}
\end{center}
\end{figure}

\begin{figure}
\begin{center}
\includegraphics[angle=0,scale=1 ]{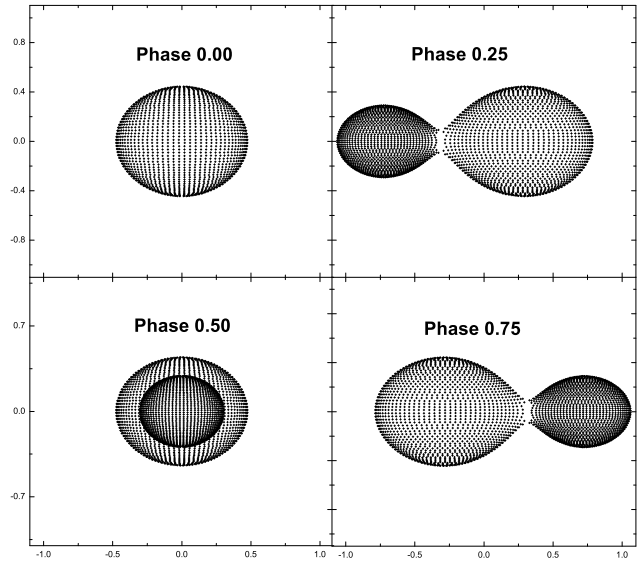}
\caption{Geometrical structure of the shallow contact binary V396
Mon at phases 0.00P, 0.25P, 0.50P and 0.75P.}
\end{center}
\end{figure}

\end{document}